\documentclass[aip,amsmath,amssymb,reprint,]{revtex4-1}

\usepackage{graphicx}
\usepackage{dcolumn}
\usepackage{bm}
\usepackage[utf8]{inputenc}
\usepackage[T1]{fontenc}
\usepackage{mathptmx}
\usepackage{etoolbox}
\usepackage{lmodern}
\usepackage{amsmath}
\usepackage{amssymb}
\usepackage{color}

\makeatletter
\def\@email#1#2{%
 \endgroup
 \patchcmd{\titleblock@produce}
  {\frontmatter@RRAPformat}
  {\frontmatter@RRAPformat{\produce@RRAP{*#1\href{mailto:#2}{#2}}}\frontmatter@RRAPformat}
  {}{}
}%
\makeatother
\begin{document}

\preprint{AIP/123-QED}

\title[Plasmon-enhanced circular dichroism spectroscopy of chiral drug solutions]{Plasmon-enhanced circular dichroism spectroscopy of chiral drug solutions}

\author{Matteo Venturi}
\affiliation{Department of Physical and Chemical Sciences, University of L'Aquila, Via Vetoio, 67100 L'Aquila, Italy}

\author{Raju Adhikary}%
\affiliation{Department of Physical and Chemical Sciences, University of L'Aquila, Via Vetoio, 67100 L'Aquila, Italy}

\author{Ambaresh Sahoo}
\affiliation{Department of Physical and Chemical Sciences, University of L'Aquila, Via Vetoio, 67100 L'Aquila, Italy}

\author{Carino Ferrante}
\affiliation{CNR-SPIN, c/o Dip.to di Scienze Fisiche e Chimiche, Via Vetoio, Coppito (L'Aquila) 67100, Italy}

\author{Isabella Daidone}
\affiliation{Department of Physical and Chemical Sciences, University of L'Aquila, Via Vetoio, 67100 L'Aquila, Italy}

\author{Francesco Di Stasio}
\affiliation{Istituto Italiano di Tecnologia, Via Morego 30, Genova 16136, Italy}

\author{Andrea Toma}
\affiliation{Istituto Italiano di Tecnologia, Via Morego 30, Genova 16136, Italy}

\author{Francesco Tani}
\affiliation{Max Planck Institute for the Science of Light, Staudtstr. 2, 91058 Erlangen, Germany} 

\author{Hatice Altug}
\affiliation{Institute of Bioengineering, Ecole Polytechnique Federale de Lausanne (EPFL), Lausanne 1015, Switzerland} 

\author{Antonio Mecozzi}
\affiliation{Department of Physical and Chemical Sciences, University of L'Aquila, Via Vetoio, 67100 L'Aquila, Italy}

\author{Massimiliano Aschi}
\affiliation{Department of Physical and Chemical Sciences, University of L'Aquila, Via Vetoio, 67100 L'Aquila, Italy}

\author{Andrea Marini}
\affiliation{Department of Physical and Chemical Sciences, University of L'Aquila, Via Vetoio, 67100 L'Aquila, Italy}
\affiliation{CNR-SPIN, c/o Dip.to di Scienze Fisiche e Chimiche, Via Vetoio, Coppito (L'Aquila) 67100, Italy}

\email{andrea.marini@univaq.it.}

\date{\today}

\begin{abstract}
We investigate the potential of surface plasmon polaritons at noble metal interfaces for surface-enhanced chiroptical sensing
of dilute chiral drug solutions. The high quality factor of surface plasmon resonances in both Otto and 
Kretschmann configurations enables the enhancement of circular dichroism differenatial absorption thanks to the 
large near-field intensity of such 
plasmonic excitations. Furthermore, the subwavelength confinement of surface plasmon polaritons is key to attain chiroptical 
sensitivity to small amounts of drug volumes placed around $\simeq 100$ nm by the metal surface. Our calculations focus on 
reparixin, a pharmaceutical molecule currently used in clinical studies for patients with community-acquired pneumonia,  
including COVID-19 and acute respiratory distress syndrome. Considering realistic dilute solutions of reparixin dissolved in 
water with concentration $\leq 5$ mg$/$ml, we find a circular-dichroism differential absorption enhancement factor 
of the order $\simeq 20$ and chirality-induced polarization distortion upon surface plasmon polariton excitation.
\end{abstract}

\maketitle

\section{\label{sec:level1}Introduction}
Chiral sensing plays a key role in pharmaceutics because the specific enantiomeric form of drugs affects their functionality 
and toxicity \cite{ChiralDrugs}. Current state-of-the-art techniques capable of measuring enantiomeric excess of chiral mixtures, 
e.g., nuclear magnetic resonance \cite{NMR}, gas chromatography \cite{Chromatography} and high performance 
liquid chromatography \cite{HPLC}, provide advanced tools for drug safety, but are designed to operate with 
macroscopic drug volumes and are not suitable for real-time analysis. Photonic devices 
can potentially overcome such limitations, but technological advancement in this direction is hampered by the inherently weak 
chiroptical interaction, which requires ml volumes in order to attain sensitivity to the enantiomeric imbalance of chiral drug 
solutions through, e.g., polarimetry \cite{Polarimetry} or electronic/vibrational circular dichroism (CD) \cite{VCD,VCDD}. 
Diverse approaches are currently investigated to enhance chiroptical interaction, as in particular ultrafast and nonlinear 
chiroptical spectroscopy techniques \cite{BeaulieuNatPhys2018,CohenPRX2019,SmirnovaNatPhot2019,Smirnova2022} or the exploitation
of superchiral fields \cite{TangPRL2010,TangScience2011} that can be engineered via metasurfaces and nanophotonic structures 
\cite{MohammadiACSPhot2018,Pellegrini2018,Gilroy2019}. 
However, the actual exploitation of superchiral fields for novel chiral 
sensing schemes remains elusive, mainly because most of the experimental sensing demonstrations in literature rely on
chiral nanoplasmonic subtrates. The major issue is that when a chiral sample is placed in the near-field of such
nanostructures, the detected CD signal is contaminated by the chiroptical response of the nanostructure itself, which acts as 
a strong background noise and masks the weak chiroptical signal of the actual sample \cite{Mohammadi2019}. 
It has to be noted that chiral drugs diluted in solutions are continuously deformed by the conformational fluctuations of the molecule in interaction with the solvent at finite temperature, modulating over time their chiroptical response. Concurring electric and magnetic resonances in 
nanostructures provide enhanced CD thanks to the dual enhancement of both electric and magnetic fields \cite{ACSPhotonics2021}. 
While such schemes are promising for single molecule chiral discrimination, they are difficult to implement for extended drug 
solutions because CD enhancement is attained only at specific hot spots where electric and magnetic field lines are strong, 
parallel and $\pi/2$ shifted \cite{Mohammadi2019,ACSPhotonics2021}. Over a complementary direction, metal-based nanophotonic structures enable 
plasmon-enhanced CD spectroscopy \cite{Nesterov2016}, which has been predicted for isolated molecules \cite{Govorov2011}, while it 
has been adopted mainly to characterise chiral nanoparticles \cite{Fan2010,Slocik2011,Hu2019}. 

\begin{figure*}[t]
\begin{center}
\includegraphics[width=\textwidth]{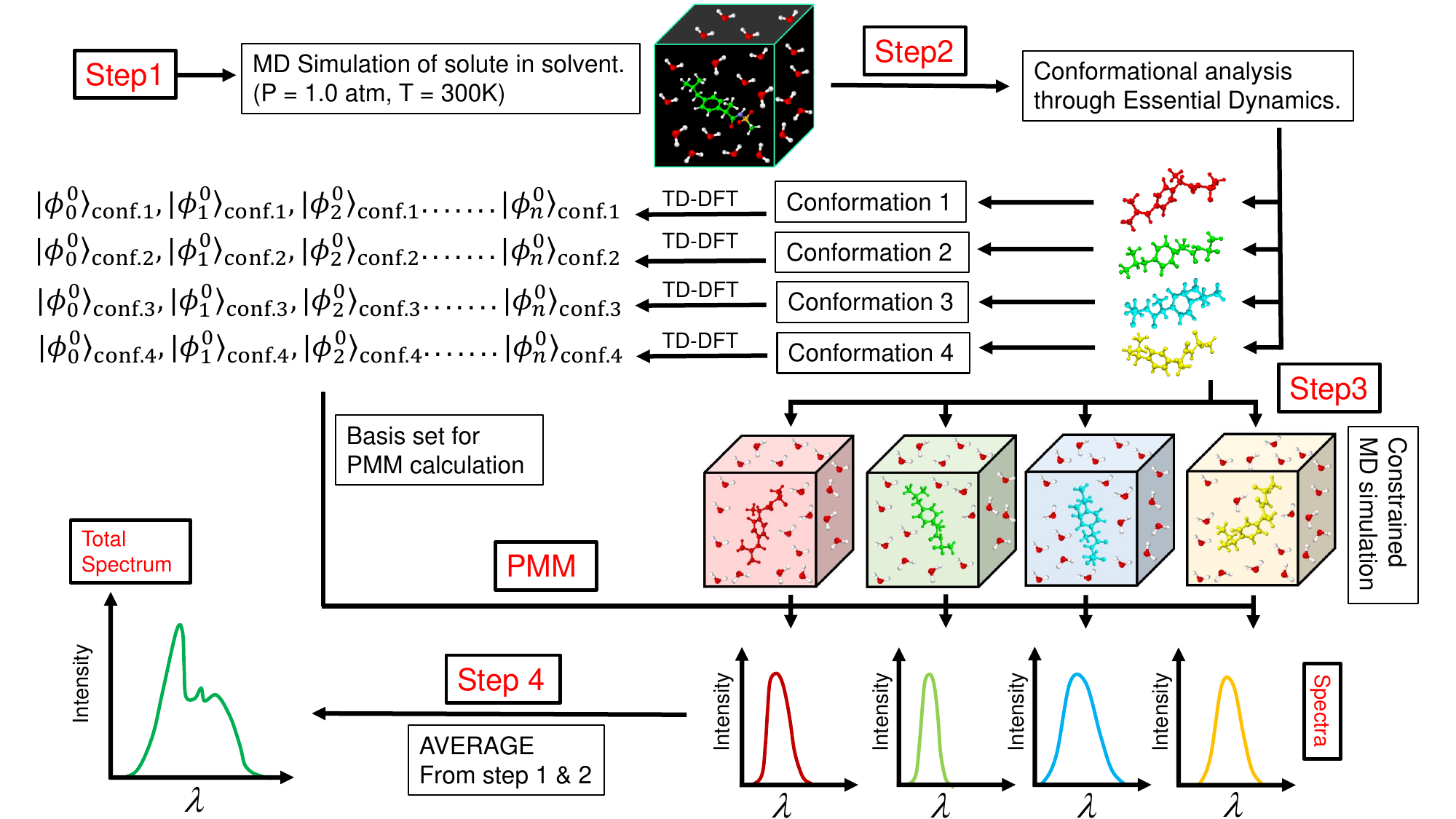}
\caption{ {\bf Step 1}. MD simulation of reparaxin in water. {\bf Step 2}. {\bf (2a)} Conformational analysis of the trajectory from {\bf Step 1} through Essential Dynamics 
(see supporting information). {\bf (2b)} Extraction of the (four) reparaxin conformation states, representative of high-probability basins 
(see, Fig. 2), for TD-DFT calculations (in the gas-phase) aimed at obtaining the unperturbed basis-set, of size $N_{\rm s}$, necessary for {\bf Step 3c}. {\bf Step 3}. {\bf (3a)} For each reparaxin conformation state a further MD simulation is carried out with solute kept frozen and the solvent allowed to move. {\bf 3b} The electric field exerted by the solvent is evaluated onto the reparaxin center of mass at each frame of the trajectory of the previous section. {\bf 3c} The basis set  obtained in the {\bf Step 2b} and the electric field from the {\bf Step 3b} Perturbed Matrix Method (see supporting information) is applied for evaluating, at each frame of the  trajectory of {\bf Step 3a}, the instantaneous expectation value of interest (electric dipole, electronic energy, vibrational energy) or the instantaneous values of electric or magnetic transition moment. The latter are used to obtain the UV and/or CD spectrum, either electronic or vibrational. {\bf Step 4}. According to the probability of conformation state basins, all the above observables (including the spectra), the observables - assuming an ergodic behaviour of the produced trajectories - are calculated as time-averages and represent ensemble averages.}
\end{center}
\end{figure*}

\begin{figure*}[t]
\begin{center}
\includegraphics[scale=1]{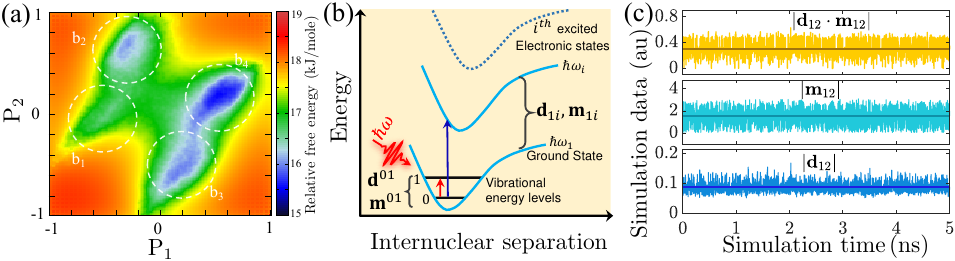}
\caption{ {\bf (a)} Calculated relative free-energy at $T = 298$ K of the reparixin pure S enantiomer (dissolved in water) through MD simulations followed by Essential Dynamics analysis (see supporting information for details), indicating the co-existence of four distinct, statistically relevant {conformation state} basins ${\rm b}_k$, where $k = 1-4$. $P_1$ and $P_2$ represent the projections of the solvated reparaxin position vectors (sampled during the MD simulation) onto the eigenvectors of the all-atoms covariance matrix (see supporting information for additional details). {\bf (b)} Sketch of reparixin vibronic dynamics produced by a driving optical laser field in the electric/magnetic dipole approximation. {\bf (c)} Temporal evolution of electronic transition dipole moduli $|{\bf d}_{1,2}|$, $|{\bf m}_{1,2}|$, and chiral projection $|{\bf d}_{1,2}\cdot{\bf m}_{1,2}|$ (in atomic units) { of a single reparixin molecule in the ${\rm b}_1$ conformation state} calculated at each temporal frame (expressed in nanoseconds) of the MD simulation through PMM calculations (see supporting information for additional details). { We emphasize that the MD simulation box size is $\simeq 3$ nm lateral size and $\simeq 27$ nm$^3$ volume, and is unrelated to the $100$ nm thickness of the chiral sample used}.}
\end{center}
\end{figure*}

Here we predict that CD differential absorption (CDDA) by dilute drug solutions can get enhanced by a factor $f_{\rm CDDA}\simeq 20$ via
the excitation of surface plasmon polaritons (SPPs) at a planar interface between a noble metal and the chiral sample. In particular, 
we focus on an isotropic assembly of reparixin (an inhibitor of the CXCR2 function attenuating inflammatory responses \cite{Bertini2007} 
that has been adopted in clinical trials for the treatment of hospitalized patients with COVID-19 pneumonia \cite{Landoni2022}) 
dissolved in water with dilute number molecular density of the order $n_{\rm mol} = n_{\rm mol}^{\rm R} + n_{\rm mol}^{\rm S} \simeq 10^{-2}$ nm$^{-3}$ (corresponding to a concentration of $5$ mg$/$ml), where $n_{\rm mol}^{\rm R,S}$ indicate the number densities of 
R and S enantiomers, respectively. Because SPPs possess a momentum greater than the one of radiation in vacuum, 
their excitation requires a radiation momentum kick, which can be provided either by a grating \cite{Park2003,Offerhaus2005} or by a 
prism through attenuated total reflection in Kretschmann \cite{Kretschmann} or Otto \cite{Otto} configurations. However, the adoption 
of gratings to excite SPPs modifies significantly the SPP dispersion \cite{Hooper2004}, and in turn here we focus on silica 
prism coupling. We observe that CDDA $\Delta A = A_{\rm R} - A_{\rm L}$ upon right/left circular polarization excitation is proportional to the enantiomeric excess density $\Delta n_{\rm mol} = n_{\rm mol}^{\rm S} - n_{\rm mol}^{\rm R}$, and 
in turn by measuring the CDDA it is possible to retrieve $\Delta n_{\rm mol}$. Furthermore, we observe that the maximum plasmon-induced
CDDA enhancement factor $f_{\rm CDDA} \simeq 20$ is attained with silver in the Otto configuration, while other noble metals like gold 
and copper provide a 
smaller $f_{\rm CDDA}$. 
The vacuum wavelength ($\lambda$) dependence of all macroscopic optical parameters [dielectric permittivity $\epsilon_{\rm r}(\lambda)$, 
magnetic permeability $\mu_{\rm r}(\lambda)$ and chiral parameter $\kappa(\lambda)$] of the isotropic chiral mixture are 
calculated from first principles. This is accomplished 
by perturbatively solving the density matrix equations accounting for the leading electronic and vibrational absorption peaks and 
by averaging the obtained polarizability tensors over the random molecular orientation through the Euler rotaton matrix approach 
\cite{Andrews2004,Valev2022}. 
We evaluate plasmon-enhanced CDDA $\Delta A$ by classical electrodynamics calculations accounting for (i) silica prism in
the Otto and Kretschmann SPP coupling schemes, (ii) the macroscopic bi-anisotropic response of the chiral mixture, and (iii) the 
dielectric response of noble metals. Our predictions of efficient plasmon-enhanced CDDA by isotropic drug solutions  
suggest novel avenues for chiroptical sensing in nanoscale environments.

For the present study, it is necessary to make use of quantum molecular observables (i.e., electric and magnetic moments, electronic and vibrational excitation energies)
evaluated for the solvated drug, i.e., aqueous reparixin in our calculations. This is not a trivial task because such quantum molecular observables should be perceived 
as genuinely quantum expectation values averaged (in statistical-mechanical terms) over a large number of reparaxin/solvent configurations. 
{ For this purpose we have adopted a combination of computational approaches, described to some extent in the Supporting Information and briefly illustrated in Fig. 1. The adopted methods combine semi-classical Molecular Dynamics (MD) simulations 
(necessary for the reparaxin-solvent conformational sampling), quantum-chemical (QM) calculations and Perturbed Matrix Method (PMM)
\cite{Aschi2001,Amadei2009,Carrillo2017} which, using the information obtained from MD simulations and QM calculations, provides the observables of interests expressed as ensemble averages. This approach is necessary for including in our model subtle but relevant effects produced by the semi-classical atomic-molecular motions (both reparaxin and water).}

\section{Electronic and vibrational structure of reparixin in water}

{ In Figs. 2a-c we illustrate in more detail the key-steps for evaluating the quantum-molecular observables of interest.
As previously described in Fig. 1 and in the supporting information, the reparixin electric/magnetic transition dipole moments and the corresponding absorption UV/vibrational spectra are evaluated as ensemble averages of solvated reparaxin in water. These high probability conformation states are representative of the free-energy basins $b_k$, where $k = 1 - 4$, located through MD simulations and ED analysis and depicted in Fig. 2 in the case of reparixin S enantiomer. We underline that, due to selection rules, we consider only two vibrational states (labelled by the index $\nu = 0, 1$) producing resonant interaction with external radiation in the mid-infrared (mid-IR) with vibrational energy $\hbar\omega_{\rm v}$ independent over the reparixin conformation state} $k$. Indeed, we found that $\hbar\omega_{\rm v}$ is practically unaffected 
by the molecular configuration {state}, i.e., the nature of the vibrational mode (calculated as described in the supporting information) 
coincides in the four conformation {states}. For every enantiomer labelled by the index $a = $ R, S, the 
radiation-unperturbed vibronic structure is approximated by independent electronic/vibrational Hamiltonians 
$\hat{\cal H}_0^{\rm el,vib}(k,a)$, given by
\begin{subequations}
\begin{align} 
& \hat{\cal H}_0^{\rm el}(k,a)  = \sum\limits_{i=1}^5 \hbar\omega_i(k) |i(k,a)\rangle\langle i(k,a)|, \\ 
& \hat{\cal H}_0^{\rm vib}(a) = \sum\limits_{\nu=0,1} (\nu + 1/2) \hbar\omega_{\rm v}  |\nu(a)\rangle\langle\nu(a)|,
\end{align}
\end{subequations} 
where $\hbar\omega_i(k)$ and $\hbar\omega_{\rm v} (\nu + 1/2)$ are the electronic/vibrational energy eigenvalues 
with eigenstates $|i(k,a)\rangle$ and $|\nu(a)\rangle$. For flexible solvated molecules such as reparixin, the effect of the 
vibronic transitions in the UV spectral shape is negligible if compared to the one produced by the chromophore and solvent conformational transitions. {In turn, the lack of fine structure in the electronic transitions is due to the effect of the 
semi-classical molecular vibrations and solvent perturbations producing a spectral broadening that prevents their observation.} Furthermore, we account 
only for the electronic states $i=1-5$ because we find that the corresponding  
energy levels are the only ones producing resonant electronic absorption peaks in the ultraviolet (UV).  Fig. 2b illustrates the schematic vibronic structure of reparixin 
for a generic {molecular conformation state}. In the electric/magnetic dipole approximation, the molecule interacts with external 
radiation through the electronic/vibrational perturbing Hamiltonians given by 
the expressions
\begin{subequations}
\begin{align}
& \hat{\cal H}_{\rm I,el}^{(k,a)}({\bf r},t) = - {\bf E}({\bf r},t)\cdot\hat{\bf d}_{\rm el}(k,a)- {\bf B}({\bf r},t)\cdot\hat{\bf m}_{\rm el}(k,a), \raisetag{3.5ex} \\
& \hat{\cal H}_{\rm I,vib}^{(a)}({\bf r},t) = - {\bf E}({\bf r},t)\cdot\hat{\bf d}_{\rm vib}(a)- {\bf B}({\bf r},t)\cdot\hat{\bf m}_{\rm vib}(a),
\end{align}
\end{subequations}
where ${\bf E}({\bf r},t)$, ${\bf B}({\bf r},t)$ are the external radiation electric and magnetic induction fields and
\begin{subequations}
\begin{align}
& \hat{\bf d}_{\rm el}(k,a)  = \sum\limits_{i,j=1}^5 {\bf d}_{i,j}(k,a)|i(k,a)\rangle\langle j(k,a)|, \\ 
& \hat{\bf m}_{\rm el}(k,a)  = \sum\limits_{i,j=1}^5 {\bf m}_{i,j}(k,a)|i(k,a)\rangle\langle j(k,a)|, \\
& \hat{\bf d}_{\rm vib}(a)   = \sum\limits_{\nu,\nu' = 0,1} {\bf d}^{\nu,\nu'}(a)|\nu(a)\rangle\langle\nu'(a)|, \\
& \hat{\bf m}_{\rm vib}(a)   = \sum\limits_{\nu,\nu' = 0,1} {\bf m}^{\nu,\nu'}(a)|\nu(a)\rangle\langle\nu'(a)|
\end{align}
\end{subequations}
indicate the dyadic representation of electric/magnetic electronic/vibrational dipole operators in the set of radiation-unperturbed 
electronic/vibrational energy states. We emphasize that, for the particular case of reparixin, due to the 
co-presence of two electrons with opposite spins in the ground state, the static magnetic dipole moments are vanishing for every 
electron eigenstate (in the Born-Oppenheimer approximation), while static electric dipole moments are finite. In turn, the 
calculation of vibrational magnetic transition dipole moments ${\bf m}^{01}$ requires going beyond the 
Born-Oppenheimer approximation, see supporting information for additional details. Note that transition electric dipole moments are 
real polar vectors ${\bf u}(k,a)\in\Re^3$, 
while transition magnetic dipole moments are purely imaginary axial vectors ${\bf w}(k,a)\in\Im^3$,
where we indicate with ${\bf u}(k,a)$ either ${\bf d}_{1i}(k,a)$ or ${\bf d}^{01}(a)$, and with
${\bf w}(k,a)$ either ${\bf m}_{1i}(k,a)$ or ${\bf m}^{01}(a)$, depending over the
electronic/vibrational dynamics considered. We define the reflection operator 
${\cal R}_{\hat{\bf n}} = \mathbb{I} - 2\hat{\bf n}\hat{\bf n}$ by the plane perpendicular to the unit vector 
$\hat{\bf n}$ where reflection symmetry is broken. Transition dipole moments of opposite enantiomers can be 
calculated by applying the reflection operator: ${\bf u}(k,{\rm S}) = {\cal R}_{\hat{\bf n}}{\bf u}(k,{\rm R})$,
${\bf w}(k,{\rm S}) = - {\cal R}_{\hat{\bf n}}{\bf w}(k,{\rm R})$. In turn, while the modulus of 
the transition dipole moments is unaffected by the enantiomeric type $|{\bf u}(k,{\rm S})| = |{\bf u}(k,{\rm R})|$,
$|{\bf w}(k,{\rm S})| = |{\bf w}(k,{\rm R})|$, the scalar product 
${\bf u}(k,{\rm S})\cdot{\bf w}(k,{\rm S}) = - {\bf u}(k,{\rm R})\cdot{\bf w}(k,{\rm R})$ flips sign for opposite
enantiomeric forms. Such a quantity, which we call {\it chiral projection}, is in turn chirally sensitive as it involves the scalar product of a polar and an axial vector. In Fig. 2c we report the MD-induced temporal evolution of electronic contributions
$|{\bf d}_{12}|$, $|{\bf m}_{12}|$ and the modulus of the chiral projection $|{\bf d}_{12}\cdot{\bf m}_{12}|$ of the S reparixin enantiomer in {conformation state} $k=1$. Note that such quantities, constituting the key 
microscopic ingredients producing optical activity, see below, depend over time due to MD. For this reason, in our 
quantum-mechanical calculations of the macroscopic bi-anisotropic response we use { ensemble-averaged 
$<|{\bf d}_{12}|>$, $<|{\bf m}_{12}|>$ and $<|{\bf d}_{12}\cdot{\bf m}_{12}|>$ (calculated as time-averages over the 
$\simeq$ ns MD timescale in our ergodic assumption, see below}.   

\begin{figure*}[t]
\begin{center}
\includegraphics[width=\textwidth]{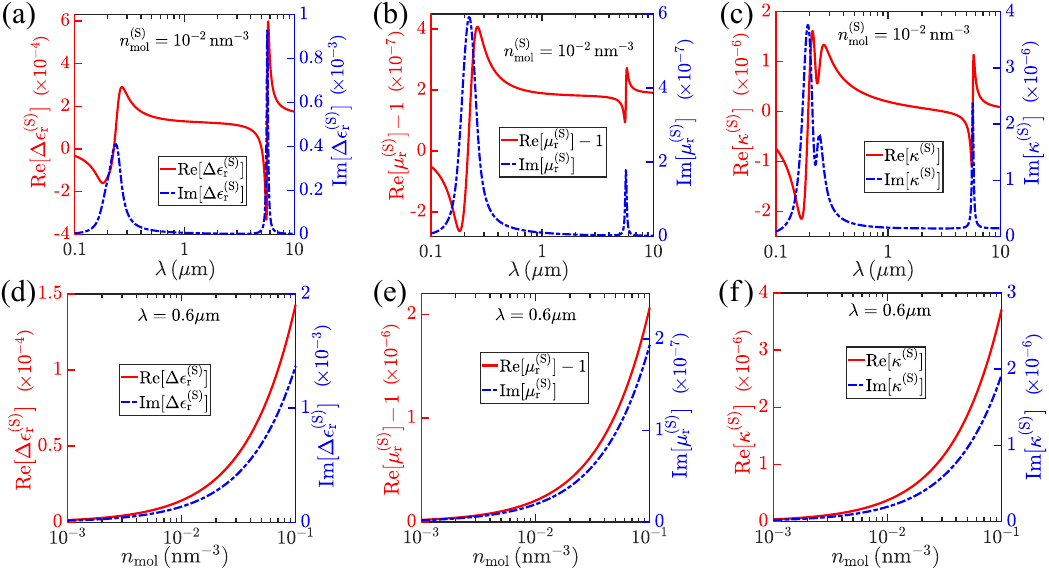}
\caption{ {\bf (a-c)} Vacuum wavelength $\lambda$ and {\bf (d-f)} number molecular density $n_{\rm mol}$ dependencies of 
{\bf (a,d)} relative dielectric permittivity correction $\Delta\epsilon_{\rm r}$, 
{\bf (b,e)} relative magnetic permeability correction $\mu_{\rm r} - 1$, and {\bf (c,f)} chiral parameter $\kappa$ 
of a water solution containing only the S enantiomer of reparixin for {\bf (a-c)} fixed molecular number density 
$n_{\rm mol} = 10^{-2}$ nm$^{-3}$ and {\bf (d-f)} fixed vacuum wavelength $\lambda = 600$ nm.}
\end{center}
\end{figure*}

\section{Macroscopic bi-anisotropic response of reparixin in water}

For every { conformation state} $k$ and enantiomeric form $a$, radiation-induced electron/vibrational dynamics of chiral drugs 
(reparixin in our calculations) is in turn governed by the density matrix equations 
\begin{equation}
\frac{d\hat\rho_{k,a}}{dt} = \frac{1}{i\hbar}\left[ \hat{\cal H}_0^{\rm el,vib}(k,a) + \hat{\cal H}_{\rm I,el,vib}^{(k,a)},\hat\rho_{k,a} \right] + \hat{\cal L}(\hat\rho_{k,a}), \label{DMEqs}
\end{equation}
where $\hat\rho_{k,a}=\hat\rho_{k,a}^{\rm el,vib}({\bf r},t)$ is the time-dependent electronic/vibrational density matrix and $\hat{\cal L}(\hat\rho_{k,a}^{\rm el,vib})$ is the Lindblad operator accounting for interaction with 
the thermal bath. The Lindblad relaxation rates are obtained by fitting the bandwidth of the absorption spectra, see supporting information for further details. For monochromatic external radiation fields 
${\bf V}({\bf r},t) = {\rm Re} \left[ {\bf V}_0({\bf r}) e^{-i\omega t}\right]$ with 
carrier vacuum wavelength $\lambda = 2\pi c/\omega$, where $c$ is the speed of light in vacuum and ${\bf V} = {\bf E},{\bf B}$, 
respectively, Eq. (\ref{DMEqs}) is solved perturbatively at first-order yielding $\hat\rho_{k,a}^{\rm el,vib}({\bf r},t)$ 
\cite{Govorov2012}. The expectation value of the induced electric/magnetic dipole moments { expressed as an ensemble average considering all the reparaxin conformation states for each enantiomer form $a$ (see Fig. 1)}, is in turn given by 
${\bf o}^{\rm el,vib}_{a}=\sum\limits_{k=1}^4 p(k) {\rm Tr}\left[\hat\rho_{k,a}^{\rm el,vib}\hat{\bf o}(k,a)\right]$, 
where $p(k) = 0.25$, see 
Fig. 2a, and ${\bf o} = {\bf d},{\bf m}$, respectively. Owing to the random molecular orientation, the induced electric/magnetic dipole moments averaged over arbitrary rotations through the Euler rotation matrix approach \cite{Andrews2004,Valev2022} are given by the expressions
\begin{subequations}
\begin{align}
& \langle{\bf d}_{a}\rangle = \epsilon_0 {\rm Re} \left\{ \left[ \alpha_{\rm e} {\bf E}_0 + c \alpha_{\rm m}^{(a)} {\bf B}_0\right]e^{-i\omega t } \right\}, \\
& \langle{\bf m}_{a}\rangle = \epsilon_0 c {\rm Re} \left\{ \left[ - \alpha_{\rm m}^{(a)} {\bf E}_0 + c \alpha_{\rm b} {\bf B}_0\right]e^{-i\omega t } \right\},  
\end{align}
\end{subequations}
where $\epsilon_0$, $\mu_0$ are the vacuum permittivity and permeability, respectively, and $\alpha_{\rm e}(\omega)$, 
$\alpha_{\rm b}(\omega)$, $\alpha_{\rm m}^{(a)}(\omega)$ are the linear isotropic electric, magnetic and mixing 
polarizabilities of every molecular enantiomer, given by the expressions 
\begin{eqnarray}
\alpha_{\rm e,b,m}^{(a)}(\omega) & = & {\cal C}_{\rm e,b,m}^{\rm (vib)}(a){\cal F}^{\rm (vib)}(\omega) +  \\
& + &\sum\limits_{k=1}^4 p(k)\sum\limits_{j=2}^{5} {\cal C}_{j,{\rm e,b,m}}^{\rm (el)}(k,a){\cal F}_{j,k}^{\rm (el)}(\omega) . \nonumber
\end{eqnarray}
In the expression above $p(k)=0.25$, ${\cal C}_{\rm e}^{\rm (vib, el)} = 2\langle|{\cal D}^{\rm (vib, el)}|^2\rangle/3\hbar$, 
${\cal C}_{\rm b}^{\rm (vib, el)} = 2\langle|{\cal M}^{\rm (vib, el)}|^2\rangle/3\hbar$ and 
${\cal C}_{\rm m}^{\rm (vib, el)} = 2i \langle{\rm Im} [{\cal D}^{\rm (vib, el)}\cdot{\cal M}^{\rm (vib, el)}]\rangle/3\hbar$, where $\langle ( ... ) \rangle$ indicates the { ensemble-average (calculated as time-average over the $\simeq$ ns MD time-scale in our ergodic assumption)}, 
${\cal D}^{\rm (vib)}={\bf d}^{01}$, ${\cal D}^{\rm (el)}_j={\bf d}_{1j}$, 
${\cal M}^{\rm (vib)}={\bf m}^{01}$ and ${\cal M}^{\rm (el)}_j={\bf m}_{1j}$. The angular frequency dependence of the polarizabilities
is accounted by ${\cal F}^{\rm (vib)}(\omega) = \displaystyle\sum\limits_{\sigma = \pm 1} \sigma/(\omega + \sigma \langle \omega_{\rm v} \rangle + i\gamma_{\rm v}/2)$ and 
${\cal F}_{j,k}^{\rm (el)}(\omega) = \displaystyle\sum\limits_{\sigma = \pm 1} \sigma/[\omega + \sigma \langle \omega_j(k) - \omega_1(k) \rangle + i\gamma_j^{\rm (e)}(k)/2]$,
where $\gamma_{\rm v} = 0.012$ fs$^{-1}$ is the relaxation rate of the only reparixin vibrational transition 
$\nu = 0 \rightarrow 1$ 
with energy $\hbar\omega_{\rm v} = 0.218$ eV (corresponding to the transition wavelength 
$\lambda = 5.69$ $\mu$m). The calculated electric/magnetic transition dipole moments 
${\bf d}^{01},{\bf m}^{01}$ 
provide $\sqrt{\langle|{\bf d}^{01}|^2\rangle}/d_0 = 0.105$, $\sqrt{\langle|{\bf m}^{01}|^2\rangle}/m_0 = 0.398$ and $\langle {\rm Im} [{\bf d}^{01}({\rm S})\cdot{\bf d}^{01}({\rm S})]\rangle/d_0m_0 = 0.007$, where $d_0 = |e| a_0$, $a_0 = \hbar/m_{\rm e}c\alpha$ is the Bohr radius, $m_0 = |e|\hbar/2m_{\rm e}$ is the Bohr magneton, 
$\alpha$ is the fine-structure constant and $e,m_{\rm e}$ are the electron charge and mass. 
The electronic relaxation rates $\gamma_j^{\rm (e)}$ have been extracted by the frequency bandwidth of the { UV absorption spectrum and are reported in Tab. S1 along with the electronic transition energies, the electric/magnetic transition
dipole moment moduli and the chiral projections expressed as ensemble averages.} While in principle such a procedure might provide inconsistent results, we find that averaging over time the full microscopic polarizabilities {(we calculate ensemble averages as 
time averages in our ergodic assumptions)} or adopting individually averaged electric/magnetic transition dipole moduli, chiral projections and transition energies provides consistent results, see Fig. S1 for comparison.  

\begin{figure*}[t]
\begin{center}
\includegraphics[width=\textwidth]{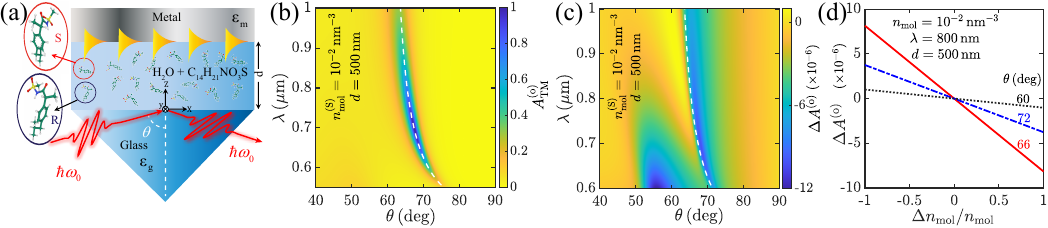}
\caption{{\bf (a)} Schematic representation of plasmon-enhanced CD in the Otto configuration. The considered chiral 
sample is a solution of	S and R reparixin enantiomers dissolved in water with molecular density $n_{\rm mol}$. 
{\bf (b,c)} Dependence of {\bf (b)} TM absorbance $A_{\rm TM}^{\rm (o)}(\theta,\lambda)$ (for TM linear polarization 
excitation) and {\bf (c)} CDDA 
$\Delta A^{\rm (o)}(\theta,\lambda) = A_{\rm R}(\theta,\lambda) - A_{\rm L}(\theta,\lambda)$ (for circular polarization 
excitations) over the vacuum wavelength $\lambda$ and the angle of incidence $\theta$ of the impinging radiation in the Otto coupling scheme illustrated in (a). The plots refer to a pure S reparixin enantiomer dissolved in water (with 
enantiomeric number density imbalance $\Delta n_{\rm mol} = n_{\rm mol}^{\rm (S)} - n_{\rm mol}^{\rm (R)} = n_{\rm mol}$). {\bf (d)} Dependence of CDDA over the enantiomeric number density imbalance
$\Delta n_{\rm mol} = n_{\rm mol}^{\rm (S)} - n_{\rm mol}^{\rm (R)}$ for several distinct incidence angles $\theta$ 
 and fixed excitation vacuum wavelength $\lambda = 800$ nm. 
All plots are obtained for a silver substrate with dielectric constant $\epsilon_{\rm m}(\lambda)$ (taken from 
Ref. \cite{SilverRef}), fixed solvate aqueous reparixin thickness $d = 500$ nm and fixed total molecular number density 
$n_{\rm mol} = 10^{-2}$ nm$^{-3}$. The dashed white curves in {\bf (b,c)} indicate the SPP dispersion relation.}
\end{center}
\end{figure*}

In the limit of dilute drug solution, { reparixin dissolved in water can be treated as an effective-medium with bi-anisotropic macroscopic response} by introducing the  
polarization 
${\bf P}({\bf r},t) = \sum\limits_{a = {\rm R,S}} n_{\rm mol}^{(a)}\langle{\bf d}_{a}({\bf r},t)\rangle$ and magnetization 
${\bf M}({\bf r},t) = \sum\limits_{a = {\rm R,S}} n_{\rm mol}^{(a)}\langle{\bf m}_{a}({\bf r},t)\rangle$ fields. Such vectorial fields depend over the orientation-averaged \cite{Andrews2004,Valev2022} electric/magnetic induced dipole moments 
$\langle{\bf d}_{a}\rangle$, $\langle{\bf m}_{a}\rangle$
and the number molecular densities $n_{\rm mol}^{\rm (R,S)}$ of the two enantiomeric forms $a={\rm R,S}$. 
In turn, the macroscopic displacement vector ${\bf D}({\bf r},t) = \epsilon_0 {\bf E}({\bf r},t) + {\bf P}({\bf r},t)$ and 
magnetic field ${\bf H}({\bf r},t) = {\bf B}({\bf r},t)/\mu_0 - {\bf M}({\bf r},t)$ are given by the 
bi-anisotropic dispersion relations
\begin{subequations}
\begin{align}
& {\bf D} = \epsilon_0{\rm Re} \left\{ \left[ \epsilon_{\rm r}(\lambda) {\bf E}_0({\bf r}) - i \mu_0 c \kappa(\lambda) {\bf H}_0({\bf r}) \right] e^{-i\omega t} \right\}, \label{BiAn1} \\
& {\bf B} = \mu_0 {\rm Re} \left\{ \left[ \frac{i\kappa(\lambda)}{\mu_0c} {\bf E}_0({\bf r}) + \mu_{\rm r}(\lambda) {\bf H}_0({\bf r}) \right] e^{-i\omega t} \right\}, \label{BiAn2}
\end{align}
\end{subequations}
where $\epsilon_{\rm r}(\lambda) = \epsilon_{\rm H_2O}(\lambda) + n_{\rm mol} \alpha_{\rm e}(\lambda)/\epsilon_0$ is the relative dielectric permittivity, $\kappa(\lambda) = i \sum\limits_{a = {\rm R,S}} n_{\rm mol}^{(a)} \alpha_{\rm m}^{(a)}(\lambda)/\epsilon_0c $ is the chiral parameter and 
$\mu_{\rm r}(\lambda) = 1 + \mu_0 n_{\rm mol} \alpha_{\rm b}(\lambda)$ is the relative magnetic permeability of the 
chiral mixture [see supplementary information, where we report a MATLAB script for the calculation of 
$\epsilon_{\rm r},\mu_{\rm r}$ and $\kappa$]. 

Because the mixing polarizability flips sign for opposite enantiomers,  
 $\alpha_{\rm m}^{\rm (R)} = - \alpha_{\rm m}^{\rm (S)}$, the chiral parameter is proportional to the enantiomeric number density
imbalance $\Delta n_{\rm mol} =  n_{\rm mol}^{\rm (S)} - n_{\rm mol}^{\rm (R)}$. Hence, optical activity vanishes for racemic mixtures  
where $n_{\rm mol}^{\rm (R)}=n_{\rm mol}^{\rm (S)}$. Note that 
$\epsilon_{\rm r}(\lambda) = \epsilon_{\rm H_2O}(\lambda) + \Delta\epsilon_{\rm r}(\lambda)$ accounts for both the dielectric permittivity of the solvent (water) $\epsilon_{\rm H_2O}(\lambda)$ \cite{WaterRef} and the calculated correction produced by 
reparixin $\Delta\epsilon_{\rm r}(\lambda)$.   
In Fig. 3a-f we illustrate the dependence of $\Delta\epsilon_{\rm r}(\lambda)$, $\kappa(\lambda)$ and $\mu_{\rm r}(\lambda)$ of
a solution of pure S reparixin enantiomers dissolved in water ($n_{\rm mol}^{\rm (R)} = 0$) over the molecular number density 
$n_{\rm mol}$ and the vacuum wavelength of 
impinging radiation $\lambda$. Note that the chiroptical response of reparixin is affected by the tails
of electronic (resonant at $\lambda\simeq 250$ nm) and vibrational transitions (resonant at $\lambda\simeq 6\mu$m). 
The real part of 
$\epsilon_{\rm r}(\lambda)$ is the leading contribution to the refractive index dispersion $n(\lambda) = {\rm Re} \left[\sqrt{\epsilon_{\rm r}(\lambda)}\right]$, which is dominated by the refractive index of water due to the dilute molecular number density 
$10^{-3}$ nm$^{-3}< n_{\rm mol} < 10^{-1}$ nm$^{-3}$, see Figs. 3a,d. The chirally insensitive extinction coefficient 
$k_{\rm ext}(\lambda) = {\rm Im} \left[\sqrt{\epsilon_{\rm r}(\lambda)}\right]$ is affected mainly by the imaginary part of 
$\epsilon_{\rm r}(\lambda)$ and is also dominated by the absorption of water, see Figs. 3a,d. The relative magnetic permeability 
$\mu_{\rm r}(\lambda)$ arises from the reparixin magnetic response but is chirally insensitive and provides only a tiny correction
to the mixture absorption and dispersion, see Figs. 3b,e. The dependence of the complex dimensionless chiral parameter 
$\kappa(\lambda)$ over vacuum wavelength $\lambda$ and molecular number density $n_{\rm mol}$ is depicted in Figs. 3c,f. Note that 
the chiral parameter is responsible for the rotatory power (proportional to ${\rm Re} [ \kappa(\lambda) ]$) and CD 
(proportional to ${\rm Im} [ \kappa(\lambda) ]$), and is maximised at the electronic/vibrational resonances of reparixin.

\section{Plasmon-enhanced CD spectroscopy}

Fig. 4a illustrates the considered Otto coupling scheme composed of a silica prism and a noble metal substrate 
separated by a distance $d$ ($100$nm$<d<1\mu$m) embedding the chiral drug solution to be analysed, reparixin dissolved 
in water in our calculations. We consider monochromatic radiation with vacuum wavelength 
$\lambda$, incidence angle $\theta$ and arbitrary polarization given by the superposition of right 
($s=+1$) and left ($s=-1$) 
circular components, impinging from a BK$7$ silica prism with relative dielectric permittivity $\epsilon_{\rm g}(\lambda)$ \cite{SilicaRef}
($z<0$, see Fig. 4a) on a layer of reparixin dissolved in water [with bi-anisotropic response given by Eqs. (\ref{BiAn1},\ref{BiAn2})] 
placed at $0<z<d$ and a metal substrate with dielectric constant $\epsilon_{\rm m}(\lambda)$ in $z>d$. 
Because circular polarization waves are eigenfunctions of Maxwell's
equations in the considered chiral medium, we decompose the impinging wave on circular polarization unit vectors 
$\hat{\bf e}_s^{(0)}(\theta) = ( {\rm cos}\theta \hat{\bf e}_x + is \hat{\bf e}_y - {\rm sin}\theta\hat{\bf e}_z )/\sqrt{2}$ and 
investigate the scattering dynamics of circular polarization waves { by analytically solving macroscopic Maxwell's equations accounting 
for the distinct polarization and magnetization fields in every medium of the considered geometry}. We obtain

\begin{figure*}[t]
\begin{center}
\includegraphics[width=\textwidth]{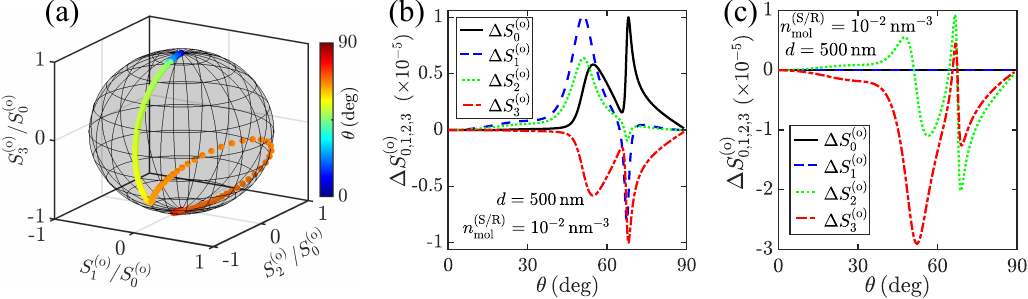}
\caption{Reflected polarization dynamics considering a silver substrate with dielectric
constant $\epsilon_{\rm m}(\lambda)$\cite{SilverRef}, vacuum wavelength $\lambda = 700$ nm, solvated aqueous reparixin with 
thickness $d= 500$ nm and total molecular number density 
$n_{\rm mol} = 10^{-2}$ nm$^{-3}$. {\bf (a)} Evolution of the Stokes parameters as a function of incidence angle $\theta$ in the 
Poincar\'e sphere for right-circular impinging polarization and pure S solvated aqueous reparixin with enantiomeric number density imbalance 
$\Delta n_{\rm mol} = n_{\rm mol}^{\rm (S)} - n_{\rm mol}^{\rm (R)} = n_{\rm mol}$. {\bf (b,c)} Stokes parameters variation between pure S
and pure R solvated aqueous reparixin upon {\bf (b)} right-circular and {\bf (c)} TM impinging polarization.} 
\end{center}
\end{figure*}

\begin{eqnarray}
& & {\bf E}_{\rm o}({\bf r},t) = \sum\limits_{s = \pm 1} {\rm Re} \left\{ \left[ {\bf E}_<^{({\rm o},s)}({\bf r}) \Theta(-z) + \right. \right.  \\
& & \left. \left. + {\bf E}_{\rm in}^{({\rm o},s)}({\bf r})\Theta_{\rm in}(z) + {\bf E}_>^{({\rm o},s)}({\bf r})\Theta(z-d) \right] e^{- i \omega t}\right\}, \nonumber \\
& & {\bf H}_{\rm o}({\bf r},t) = \sum\limits_{s = \pm 1} {\rm Re} \left\{ \frac{s}{i\mu_0c}\left[ \sqrt{\epsilon_{\rm g}} {\bf E}_<^{({\rm o},s)}({\bf r}) \Theta(-z) + \right.\right.  \\
& & \left. \left. + \frac{\sqrt{\epsilon_{\rm r}}}{\sqrt{\mu_{\rm r}}} {\bf E}_{\rm in}^{({\rm o},s)}({\bf r})  \Theta_{\rm in}(z) + \sqrt{\epsilon_{\rm m}} {\bf E}_>^{({\rm o},s)}({\bf r}) \Theta(z-d) \right] e^{- i \omega t}  \right\}, \nonumber
\end{eqnarray}
where the subscript/superscript ``${\rm o}$'' indicates the considered Otto configuration,
$\Theta(z)$ is the Heaviside step function, $\Theta_{\rm in}(z)= \Theta(z) - \Theta(z-d)$,
\begin{subequations}
\begin{align}
& {\bf E}_<^{({\rm o},s)}({\bf r}) = \left[ E_{0,s}^{\rm (o)}\hat{\bf e}_s^{(0)} e^{i k_{\rm g} z} + E_{{\rm R},s}^{\rm (o)} \hat{\bf e}_s^{\rm (R)} e^{- i k_{\rm g} z} \right] e^{i k_x x}, \\
& {\bf E}_{\rm in}^{({\rm o},s)}({\bf r}) = \sum\limits_{\sigma = \pm 1} E_{\sigma,s}^{\rm (o)} \left[ \sigma \hat{e}_x - i\frac{k_0}{\beta_s}(\kappa - s\sqrt{\epsilon_{\rm r}\mu_{\rm r}})\hat{e}_y + \right. \nonumber \\
& \left. - \frac{k_x}{\beta_s}\hat{e}_z \right] e^{i k_x x + i\sigma \beta_s z} , \\
& {\bf E}_>^{({\rm o},s)}({\bf r}) = \frac{1}{\sqrt{2}} E_{{\rm T},s}^{\rm (o)}\left[ \sqrt{1 - \frac{ \epsilon_{\rm g}}{ \epsilon_{\rm m}} {\rm sin}^2\theta} \hat{\bf e}_x + is \hat{\bf e}_y + \right. \nonumber \\
& \left. - \sqrt{ \epsilon_{\rm g}/\epsilon_{\rm m} } {\rm sin}\theta\hat{\bf e}_z \right] e^{i k_x x + i k_{\rm m} z},
\end{align}
\end{subequations}

\begin{figure*}[t]
\begin{center}
\includegraphics[width=\textwidth]{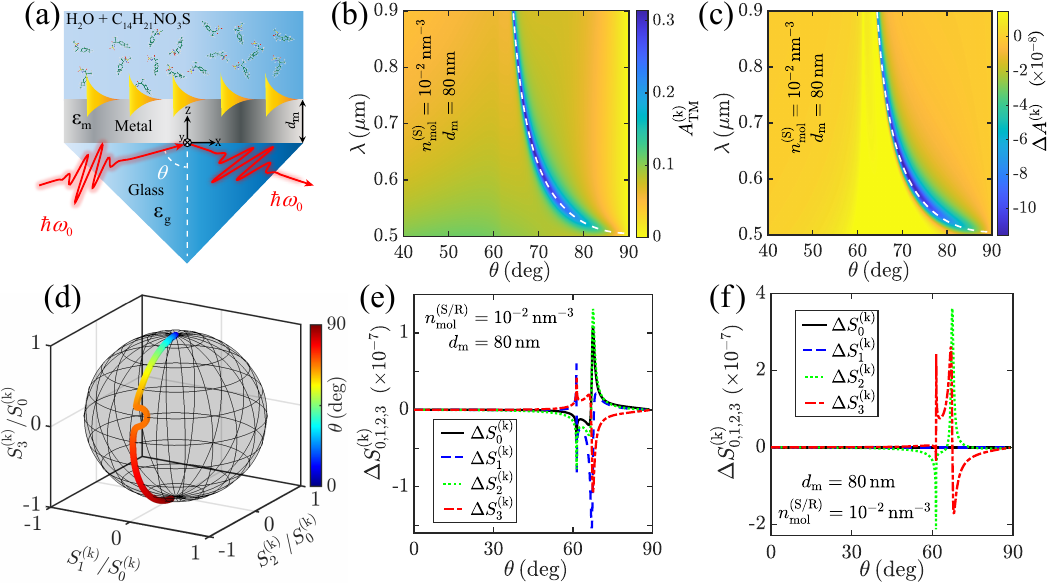}
\caption{{\bf (a)} Schematic representation of plasmon-enhanced CD in the Kretschmann configuration. The considered chiral sample is 
a solution of S and R reparixin enantiomers dissolved in water with molecular density $n_{\rm mol}$. {\bf (b,c)} Dependence of {\bf (b)} TM absorbance $A_{\rm TM}(\theta,\lambda)$ and {\bf (c)} CDDA 
$\Delta A(\theta,\lambda) = A_{\rm R}(\theta,\lambda) - A_{\rm L}(\theta,\lambda)$ (for circular polarization excitation) over the 
vacuum wavelength $\lambda$ and angle of incidence $\theta$ of the impinging radiation. Both plots are obtained for a 
silver thin film with thickness $d_{\rm m} = 80$ nm, dielectric constant $\epsilon_{\rm m}(\lambda)$ (taken from Ref. \cite{SilverRef}) and a chiral medium with total molecular number density $n_{\rm mol} = 10^{-2}$ nm$^{-3}$ and enantiomeric 
number density imbalance $\Delta n_{\rm mol} = n_{\rm mol}^{\rm (S)} - n_{\rm mol}^{\rm (R)} = n_{\rm mol}$. {\bf (d)} Evolution of 
the Stokes parameters as a function of incidence angle $\theta$ in the Poincar\'e sphere for right-circular impinging polarization, pure S solvated aqueous reparixin with enantiomeric number density imbalance  
$\Delta n_{\rm mol} = n_{\rm mol}^{\rm (S)} - n_{\rm mol}^{\rm (R)} = n_{\rm mol}$ and impinging vacuum wavelength $\lambda = 700$ nm. 
{\bf (e,f)} Stokes parameters variation between pure S
and pure R solvated aqueous reparixin upon {\bf (e)} right-circular and {\bf (f)} TM impinging polarization for impinging vacuum wavelength $\lambda = 700$ nm. 
}
\end{center}
\end{figure*}

\noindent $\omega = 2\pi c/\lambda$ is the angular frequency, $k_0 = \omega/c$, $E_{0,\pm 1}^{\rm (o)}$ 
are the projections of the impinging vectorial amplitude over $\hat{\bf e}_{\pm 1}^{(0)}$, 
$\hat{\bf e}_s^{\rm (R)} = ( - {\rm cos}\theta \hat{\bf e}_x + is \hat{\bf e}_y - {\rm sin}\theta\hat{\bf e}_z )/\sqrt{2} $, 
$k_x = k_0 \sqrt{\epsilon_{\rm g}}{\rm sin}\theta$ is the conserved impinging ${\bf k}$-vector $x$-component, 
$\beta_s = \sqrt{k_0^2 (\kappa - s \sqrt{\epsilon_{\rm r}\mu_{\rm r}})^2 - k_x^2}$ is the polarization-dependent wave-vector $z$-component 
leading to optical activity and $k_{\rm g, m} = \sqrt{k_0^2\epsilon_{\rm g,m} - k_x^2}$. In order to calculate the $s$-dependent 
forward ($\sigma = + 1$) and backward ($\sigma = - 1$) amplitudes within the chiral medium $E_{\sigma,s}^{\rm (o)}$, the reflected
($E_{{\rm R},s}^{\rm (o)}$) and transmitted ($E_{{\rm T},s}^{\rm (o)}$) field amplitudes, we apply the boundary conditions (BCs) for the 
continuity of (i) the 
normal components of the displacement vector and induction magnetic field, and 
(ii) the tangential components of the electric and magnetic fields at the interfaces $z=0,d$. 
Such BCs provide an $8\times8$ inhomogeneous system of algebraic equations for the field amplitudes, see Eq. (S1) in the supporting 
information, which we invert numerically, obtaining $E_{{\rm R},\pm 1}^{\rm (o)},E_{{\rm T},\pm 1}^{\rm (o)}$ for any impinging field components
$E_{0,\pm 1}^{\rm (o)}$. Note that, for racemic mixtures such that $n_{\rm mol}^{\rm (S)} = n_{\rm mol}^{\rm (R)}$ and $\kappa = 0$, 
electromagnetic excitations of the system, see Fig. 4a, can be split into independent transverse magnetic (TM) and transverse 
electric (TE) components. Conversely, owing to the chirality of reparixin enantiomers, non-racemic mixtures such that 
$n_{\rm mol}^{\rm (S)} \neq n_{\rm mol}^{\rm (R)}$ produce mixed polarization dynamics. The absorbance of the system upon 
right ($s=+1$, $E_{0,-}^{\rm (o)} = 0$) and left ($s=-1$, $E_{0,+}^{\rm (o)} = 0$) circular polarization excitation is in turn given by 
$A_s^{\rm (o)}(\theta,\lambda) = 1 - \sum\limits_{\sigma = \pm 1} |E_{{\rm R},\sigma}^{\rm (o)}|^2/|E_{0,s}^{\rm (o)}|^2$, from which we calculate the CDDA 
$\Delta A^{\rm (o)}(\theta,\lambda) = A_+^{\rm (o)} - A_-^{\rm (o)}$. The TM absorbance 
$A_{\rm TM}^{\rm (o)}(\theta,\lambda) = 1 - \sum\limits_{\sigma = \pm 1} |E_{{\rm R},\sigma}^{\rm (o)}|^2/|E_{0}|^2$ 
is obtained by calculating $E_{{\rm R},\sigma}^{\rm (o)}$ for TM impinging polarization, i.e., $E_{0,+}^{\rm (o)} = E_{0,-}^{\rm (o)} = E_0$. In Fig. 4b we depict 
the dependence of $A_{\rm TM}^{\rm (o)}$ over $\lambda,\theta $ considering silver as a metal substrate. Note that the absorbance 
is maximised at the SPP excitation curve 
\begin{equation}
\theta_{\rm SPP}(\lambda) \simeq {\rm arcsin} \left\{ {\rm Re}  \sqrt{ \frac{\epsilon_{\rm r}(\lambda)\epsilon_{\rm m}(\lambda) }{\epsilon_{\rm g}(\lambda) [\epsilon_{\rm g}(\lambda) + \epsilon_{\rm m}(\lambda)]} } \right\},
\end{equation}
which is indicated by the white lines in Figs. 4b,c. As a consequence of SPP resonant absorption CDDA is enhanced at the SPP dispersion curve, see Fig. 4c where we plot $\Delta A^{\rm (k)}$ as a 
function of $\lambda,\theta$. For right/left circular polarization excitation, also a Fabry-Perot resonance is excited
at $600$nm$<\lambda<800$nm and $50^\circ<\theta<60^\circ$, which enhances CDDA by a factor $f_{\rm CDDA} \simeq 10$.

\begin{figure*}[t]
\begin{center}
\includegraphics[width=\textwidth]{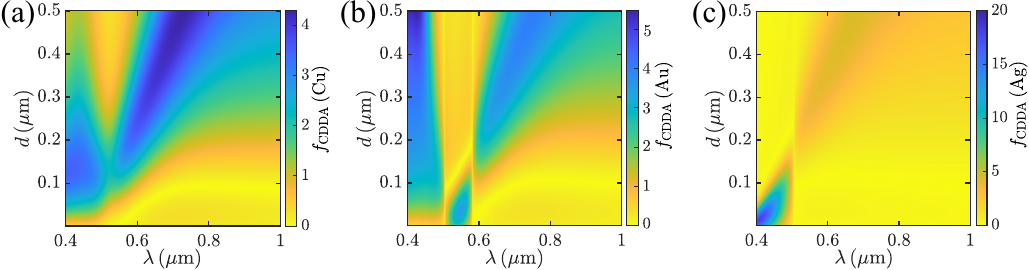}
\caption{Plasmon-induced CDDA enhancement factor $f_{\rm CDDA}(\lambda) = {\rm max}_\theta |\Delta A^{\rm (o,k)}|/|\Delta A_{\rm bulk}|$ in the Otto coupling scheme for {\bf (a)} copper, {\bf (b)} gold and {\bf (c)} silver substrates. The physical parameters used are identical to the ones used in Figs. 4a,b, except the metal type included in $\epsilon_{\rm m}$, obtained from Ref. \cite{SilverRef} for all the considered noble metals.}
\end{center}
\end{figure*} 

\section{Discussion}

Non-racemic mixtures such that $n_{\rm mol}^{\rm (S)} \neq n_{\rm mol}^{\rm (R)}$ produce mixed polarization dynamics analogously 
to optical rotation in bulk chiral media. The reflected radiation polarization can be fully tracked through
the Stokes parameters $S_0^{(\rm o)} = |E_{{\rm R},+}^{(\rm o)}|^2 + |E_{{\rm R},-}^{(\rm o)}|^2$, $S_1^{(\rm o)} = 2{\rm Re} [ E_{{\rm R},+}^{(\rm o)}E_{{\rm R},-}^{(\rm o)*} ]$, 
$S_2^{(\rm o)} = - 2{\rm Im} [ E_{{\rm R},+}^{(\rm o)}E_{{\rm R},-}^{(\rm o)*} ]$, $S_3^{(\rm o)} = |E_{{\rm R},+}^{(\rm o)}|^2 - |E_{{\rm R},-}^{(\rm o)}|^2$, whose dependence over the incidence angle $\theta$ is depicted in Figs. 5a-c at $\lambda = 700$ nm for a silver substrate. At the SPP excitation angle $\theta_{\rm SPP} \simeq 70^\circ$
polarization modulation is maximised for both circular and TM excitation polarization. Moreover, at $\theta \simeq 50^\circ $ polarization
dynamics is also enhanced due to Fabry-Perot resonance of solvated reparixin. Such a phenomenon does not occur when the drug solution is probed in the Kretschmann coupling scheme, schematically depicted in Fig. 6a.   
Here, analytical solutions of Maxwell's equations provide
\begin{eqnarray}
& & {\bf E}_{\rm k}({\bf r},t) = \sum\limits_{s = \pm 1} {\rm Re} \left\{ \left[ {\bf E}_<^{({\rm k},s)}({\bf r}) \Theta(-z) + \right. \right.  \\
& & \left. \left. + {\bf E}_{\rm in}^{({\rm k},s)}({\bf r})\Theta_{\rm in}(z) + {\bf E}_>^{({\rm k},s)}({\bf r})\Theta(z-d) \right] e^{- i \omega t}\right\}, \nonumber \\
& & {\bf H}_{\rm k}({\bf r},t) = \sum\limits_{s = \pm 1} {\rm Re} \left\{ \frac{s}{i\mu_0c}\left[ \sqrt{\epsilon_{\rm g}} {\bf E}_<^{({\rm k},s)}({\bf r}) \Theta(-z) + \right.\right.  \\
& & \left. \left. + \sqrt{\epsilon_{\rm m}} {\bf E}_{\rm in}^{({\rm k},s)}({\bf r})  \Theta_{\rm in}(z) + \frac{\sqrt{\epsilon_{\rm r}}}{\sqrt{\mu_{\rm r}}} {\bf E}_>^{({\rm k},s)}({\bf r}) \Theta(z-d) \right] e^{- i \omega t}  \right\}, \nonumber
\end{eqnarray}
where
\begin{subequations}
\begin{align}
& {\bf E}_<^{({\rm k},s)}({\bf r}) = \left[ E_{0,s}^{\rm (k)}\hat{\bf e}_s^{(0)} e^{i k_{\rm g} z} + E_{{\rm R},s}^{\rm (k)} \hat{\bf e}_s^{\rm (R)} e^{- i k_{\rm g} z} \right] e^{i k_x x}, \\
& {\bf E}_{\rm in}^{({\rm k},s)}({\bf r}) = \sum\limits_{\sigma = \pm 1} \frac{1}{\sqrt{2}} E_{\sigma,s}^{\rm (k)}\left[ \sigma \sqrt{1 - \frac{ \epsilon_{\rm g}}{ \epsilon_{\rm m}} {\rm sin}^2\theta} \hat{\bf e}_x + is \hat{\bf e}_y + \right. \nonumber \\
& \left. - \sqrt{ \epsilon_{\rm g}/\epsilon_{\rm m} } {\rm sin}\theta\hat{\bf e}_z \right] e^{i k_x x + i \sigma k_{\rm m} z}, \\
& {\bf E}_>^{({\rm k},s)}({\bf r}) =  E_{{\rm T},s}^{\rm (k)} \left[ \hat{e}_x - i\frac{k_0}{\beta_s}(\kappa - s\sqrt{\epsilon_{\rm r}\mu_{\rm r}})\hat{e}_y + \right. \nonumber \\
& \left. - \frac{k_x}{\beta_s}\hat{e}_z \right] e^{i k_x x + i \beta_s z}.
\end{align}
\end{subequations}
Note that in the expressions above the roman subscript/superscript ``${\rm k}$'' indicates the considered Kretschmann 
configuration. Again, applying BCs one gets an $8\times8$ inhomogeneous system of algebraic equations for the field amplitudes, see Eq. (S10) in 
the supporting information, which we invert numerically, obtaining $E_{{\rm R},\pm 1}^{\rm (k)},E_{{\rm T},\pm 1}^{\rm (k)}$ for any impinging field components $E_{0,\pm 1}^{\rm (k)}$. Hence, in the Kretschmann coupling scheme, the absorbance upon right 
($s=+1$, $E_{0,-}^{\rm (k)} = 0$) and left ($s=-1$, $E_{0,+}^{\rm (k)} = 0$) circular polarization excitation is given by 
$A_s^{\rm (k)}(\theta,\lambda) = 1 - \sum\limits_{\sigma = \pm 1} |E_{{\rm R},\sigma}^{\rm (k)}|^2/|E_{0,s}^{\rm (k)}|^2$, from which we calculate the Kretschmann CDDA 
$\Delta A^{\rm (k)}(\theta,\lambda) = A_+^{\rm (k)} - A_-^{\rm (k)}$. Similarly to the Otto coupling scheme, the TM absorbance 
$A_{\rm TM}^{\rm (k)}(\theta,\lambda) = 1 - \sum\limits_{\sigma = \pm 1} |E_{{\rm R},\sigma}^{\rm (k)}|^2/|E_{0}|^2$ 
is obtained by calculating $E_{{\rm R},\sigma}^{\rm (k)}$ for TM impinging polarization, i.e., by setting 
$E_{0,+}^{\rm (k)} = E_{0,-}^{\rm (k)} = E_0$. In Figs. 6b,c we depict the dependence of (b) $A_{\rm TM}^{\rm (k)}$ and 
(c) $\Delta A^{\rm (k)}$ over $\lambda,\theta $ considering a silver thin film with thickness $d_{\rm m} = 80$ nm. Note that, similarly to the Otto coupling 
scheme, the TM absorbance and the CDDA are maximised at the SPP excitation curve $\theta_{\rm SPP}(\lambda)$, indicated by the white lines in Figs. 6b,c. Conversely to the Otto coupling scheme, Fabry-Perot resonances are not excited due to the intermediate metallic film. However, polarization dependent total interal reflection between glass and aqueous reparixin produces a CDDA absorption peak at $\theta\simeq 61^\circ$. It is worth emphasizing that such a phenomenon 
occurs only thanks to the sub-wavelength metallic thickness $d_{\rm m}<<\lambda$, enabling effective evanescent coupling between glass and aqueous reparixin. Also, note that
the Kretschmann CDDA is about two orders of magnitude smaller than the Otto CDDA for identical total molecular number density 
$n_{\rm mol}$
and enantiomeric number density imbalance $\Delta n_{\rm mol}$. We illustrate the dependence of the Kretschmann Stokes parameters 
[$S_0^{(\rm k)} = |E_{{\rm R},+}^{(\rm k)}|^2 + |E_{{\rm R},-}^{(\rm k)}|^2$, 
$S_1^{(\rm k)} = 2{\rm Re} [ E_{{\rm R},+}^{(\rm k)}E_{{\rm R},-}^{(\rm k)*} ]$, 
$S_2^{(\rm k)} = - 2{\rm Im} [ E_{{\rm R},+}^{(\rm k)}E_{{\rm R},-}^{(\rm o)*} ]$, 
$S_3^{(\rm k)} = |E_{{\rm R},+}^{(\rm k)}|^2 - |E_{{\rm R},-}^{(\rm k)}|^2$ ] over the 
incidence angle $\theta$ in Figs. 6d-f at $\lambda = 700$ nm for a silver thin film with thickness $d_{\rm m} = 80$ nm and a 
chiral medium 
with $n_{\rm mol} = 10^{-2}$ nm$^{-3}$ and $\Delta n_{\rm mol} = n_{\rm mol}$. Again, note that the Stokes parameters modulation is much weaker in the Kretschmann coupling scheme, similarly to CDDA.

In order to quantify the plasmon-induced CDDA enhancement, we define the factor 
$f_{\rm CDDA}(\lambda) = {\rm max}_\theta |\Delta A^{\rm (o,k)}(\theta,\lambda)|/|\Delta A_{\rm bulk}(\lambda)|$, where $\Delta A_{\rm bulk}(\lambda) = e^{-2{\rm Im} \beta_-^{(\rm b)} d} - e^{-2{\rm Im} \beta_+^{(\rm b)} d}$ is the absorbance of a layer of reparixin dissolved in water in the absence of the nanophotonic structure and 
$\beta_\pm^{(\rm b)}= k_0 ( \sqrt{\epsilon_{\rm r}\mu_{\rm r}}\mp \kappa )$ . In Figs. 7a-c we depict the vacuum wavelength 
dependence of the CDDA enhancement for silver, gold and copper interfaces in the Otto coupling scheme. Note that 
the maximum 
plasmon-induced CDDA enhancement $f_{\rm CDDA}\simeq 20$ is attained in the Otto configuration at $\lambda\simeq 400$ nm adopting silver as metal substrate, see Fig. 7c. However, for $\lambda > 500$ nm, gold and copper provide 
CDDA enhancement factors $f_{\rm CDDA}\simeq 5$ comparable with silver. For the 
considered dilute molecular number density $n_{\rm mol} \simeq 10^{-2}$ nm$^{-3}$, we find that plasmon-enhanced CD spectroscopy is a viable chiroptical sensing platform able to discern $\lesssim 10^{13}$ ${\rm R},{\rm S}$ enantiomers in the considered $V_{\rm int}$.
{ We note that, in the considered Otto and Kretschmann coupling schemes, the transmitted intensity vanishes due to absorption because we assumed infinitely extended metal/chiral-sample for $z>d,d_{\rm m}$. However, in practical experimental realizations, the physical dimensions of the metal/chiral-sample is finite and the transmitted intensity is non-vanishing. In turn, experimentally, absorbance can be measured as $A = 1 - I_{\rm R}/I_0 - I_{\rm T}/I_0$, where $I_{\rm R,T,0}$ indicate the measured reflected ($I_{\rm R}$), transmitted 
($I_{\rm T}$), and impinging ($I_0$) intensities. CDDA can be in turn measured as the absorbance difference upon left/right impinging polarization. The dissymmetry factor can be measured as the CDDA divided by the averaged absorbance upon left and right impinging polarization. We emphasize that, while the CDDA is enhanced due to SPP excitation, the dissymmetry factor remains unaffected by the nanostructure. Indeed, the role played by SPPs lies in the enhancement of the local electric field enabling both higher averaged absorbance and higher CDDA, while maintaining their ratio (dissymmetry factor) unaffected. In turn, the proposed CD spectroscopic schemes 
are distinct from superchirality approaches  \cite{TangPRL2010,TangScience2011,MohammadiACSPhot2018,Pellegrini2018,Gilroy2019} aiming at enhancing the dissymmetry factor . Nevertheless, importantly the enhanced CDDA enables reducing the volume of the drug solution to be analysed, as illustrated above.}

\section{Conclusions}

We have investigated the potential of SPPs at interfaces between noble metals and a water solution of reparixin 
for plasmon-enhanced CD spectroscopy. The considered drug solution is composed of a realistic molecule
adopted in clinical studies for patients with community-acquired pneumonia, including COVID-19 and acute respiratory distress 
syndrome. Our calculations of reparixin microscopic observables are based on MD simulations, TD-DFT and PMM simulations enabling 
the evaluation of time-averaged permanent and transition electric/magnetic dipole moments. Macroscopic optical parameters 
of the isotropic chiral mixture are calculated by perturbatively solving the density matrix equations. Our results indicate that SPP
excitation enables plasmon-induced CDDA enhancement of the order $f_{\rm CDDA}\simeq 20$.

\section*{Supporting information}

See the supporting information for the theoretical details on the reparixin quantum observables calculations, 
the  system of algebraic equations for field amplitudes in the Otto and Kretschmann configurations, and a MATLAB script for the 
calculation of the macroscopic bi-anisotropic response of reparixin dissolved in water.

\begin{acknowledgments}
This work has been partially funded by the European Union - NextGenerationEU under the Italian Ministry of 
University and Research (MUR) National Innovation Ecosystem grant ECS00000041 - VITALITY - CUP E13C22001060006, the 
Progetti di ricerca di Rilevante Interesse Nazionale (PRIN) of the Italian Ministry of Research PHOTO (PHOtonics Terahertz devices based on tOpological materials) 2020RPEPNH, and the European Union under grant agreement No 101046424. Views and opinions expressed are however those of the author(s) only and do not necessarily reflect those of the European Union or the European Innovation Council. Neither the European Union nor the European Innovation Council can be held responsible for them. \\
The authors acknowledge fruitful discussions with Jens Biegert, Patrice Genevet, Michele Dipalo, Giovanni Melle, Sotirios Christodoulou,
Anna Maria Cimini and Domenico Bonanni.
\end{acknowledgments}

\section*{Author declarations}

\subsection*{Conflict of Interest}

The authors have no conflicts to disclose.

\section*{Author Contributions}

{\bf Matteo Venturi}: Data curation (equal); Formal analysis (equal); Methodology (supporting); Investigation (equal); Software (supporting); Visualization (supporting); Writing - original draft (equal); Writing - review and editing (equal). {\bf Raju Adhikary}: Data curation (equal); Formal analysis (equal); Methodology (supporting); Investigation (equal); Software (supporting); Visualization (supporting); Writing - original draft (supporting); Writing - review and editing (equal). {\bf Ambaresh Sahoo}: Data curation (supporting); Formal analysis (equal); Methodology (supporting); Investigation (equal); Software (supporting); Visualization; Writing - original draft (supporting); Writing - review and editing (equal). {\bf Carino Ferrante}: Methodology (supporting); Investigation (supporting); Project administration (supporting); Supervision (supporting); Visualization (supporting); Writing - original draft (supporting); Writing - review and editing (equal). {\bf Isabella Daidone}: Formal analysis (supporting); Methodology (supporting); Investigation (supporting); Software (supporting); Resources (supporting); Supervision (supporting); Writing - original draft (supporting); Writing - review and editing (equal). {\bf Francesco Di Stasio}: Conceptualization (supporting); Funding acquisition (equal); Project administration (supporting);  Writing - review and editing (equal). {\bf Andrea Toma}: Conceptualization (supporting); Funding acquisition (equal); Project administration (supporting);  Writing - review and editing (equal). {\bf Francesco Tani}: Conceptualization (supporting); Funding acquisition (equal); Project administration (supporting);  Writing - review and editing (equal). {\bf Hatice Altug}: Conceptualization (supporting); Funding acquisition (equal); Project administration (supporting);  Writing - review and editing (equal). {\bf Antonio Mecozzi}: Formal analysis (supporting); Methodology (supporting); Investigation (supporting); Writing - review and editing (equal). {\bf Massimiliano Aschi}: Data curation (supporting); Formal analysis (supporting); Methodology (equal); Investigation (supporting); Software; Resources; Supervision (equal); Writing - original draft (supporting); Writing - review and editing (equal). {\bf Andrea Marini}: Conceptualization; Formal analysis (equal); Funding acquisition (equal); Investigation (equal); Methodology (equal); Project administration; Software (supporting); Supervision (equal); Visualization (supporting); Writing - original draft (equal); Writing - review and editing (equal).

\section*{Data Availability Statement}

The data that support the findings of this study are available within the article [and its supplementary material].

\end{document}